\documentclass[12pt,a4paper]{article}
  \usepackage{a4wide}
  \usepackage{latexsym}
  \usepackage{epsf}
  \usepackage{amssymb}
  \usepackage{graphicx}
  \usepackage{amsmath, cite}
  \usepackage{amsmath,amssymb,amsthm}
  \usepackage{verbatim}
  \usepackage{hyperref}
\renewcommand{\d}{\textrm{d}}
\newcommand{\e}{\textrm{e}}

\newcommand{\be}{\begin{equation}}
\newcommand{\ee}{\end{equation}}

\begin{document}
\numberwithin{equation}{section}

\vspace{0.4cm}
\begin{center}

{\LARGE \bf{A comment on no-force conditions \\ \vspace{0.3cm} for black holes and branes}}

\vspace{2 cm} {\large Thomas Van Riet}\\
 \vspace{1 cm} {\small\slshape Instituut voor Theoretische Fysica, K.U.Leuven, \\Celestijnenlaan 200D, B-3001 Leuven, Belgium\\
 	and\\
 Institutionen f\"{o}r Fysik och Astronomi,\\ Box 803, SE-751 08 Uppsala, Sweden}
 
\vspace{.8cm} {\upshape\ttfamily  Thomas.VanRiet \emph{at} kuleuven.be }\\


\vspace{3cm}

{\bf Abstract} \end{center} {\small In the context of the Weak Gravity Conjecture the notion of quasi-extremality for black holes and branes was recently defined as the property of having either vanishing horizon size or surface gravity. It was derived that such objects obey a no-force condition. In this short note I present a simplified derivation that is essentially present in the formalism of timelike reduction pioneered by Breitenlohner, Gibbons and Maison. This formalism also provides the natural definition of quasi-extremality for gravitational instantons (and wormholes) sourced by axion fluxes and strengthens the argument that macroscopic axion wormholes do not contribute in the path integral since they are self-repulsive in a Euclidean sense.}
\newpage

\section{Weak gravity and repulsive forces} 
One of the main goals of string theory (or any other attempt to UV complete field theories coupled to gravity) is to constrain the set of possible low energy effective field theories. This endeavor has seen two approaches. Historically most work was carried out by direct computation; one would consider critical string theories and compactify them to lower dimensions and investigate the effective field theory. The typical attitude was focused on trying to reproduce appealing effective field theories: extensions of the standard model with a dark sector and hopefully a positive cosmological constant. This seems however much to ask for and perhaps string theory is not yet sufficiently developed to provide us with enough tools to be exhaustive and precise in this undertaking. Luckily the advent of the Swampland program \cite{Vafa:2005ui}\footnote{For reviews see \cite{Brennan:2017rbf, Palti:2019pca}} has brought some change. Instead of trying one compactification or another, the focus is on finding patterns and above all, understand why these patterns occur. The focus then lies on a mixture of heuristics generated by gedanken experiments (typically involving black holes), precise results from compactifications and holography. The hope is that eventually we increase our understanding of the constraints put by UV completion.  

One such constraint is given by the Weak Gravity Conjecture (WGC) \cite{ArkaniHamed:2006dz}. It roughly states that there is always some charged particle in the spectrum of the theory whose charge-to-mass ratio is at least as large as that of an extremal black hole. But supersymmetric string compactifications tend to come with massless scalars and they could give extra long range forces. This has led to the formulation of a related conjecture, called the Repulsive Force Conjecture (RFC) \cite{Palti:2017elp}, see also \cite{Lust:2017wrl, Heidenreich:2019zkl, Gendler:2020dfp}. This conjecture roughly states that there is a charged particle with the property that two copies of the particle repel each other when they are far apart\footnote{The RFC also appeared in the original paper \cite{ArkaniHamed:2006dz} to some extend.}.
It could very well be that both conjectures are true. 

In this context, where massless scalars are present, it is interesting to look at black holes (and branes) that are neither self-repulsive nor attractive. These are black holes that obey a force cancellation condition: when two such black holes are put some distance from each other, they stay put. In a recent paper \cite{Heidenreich:2020upe} Heidenreich nicely explained in detail how the vanishing of either the surface gravity or the horizon area implies such a no-force condition for general theories of gravity coupled to Abelian gauge fields and massless scalars. The main point of this small note is to explain that this result was in a sense known in the older literature related to \cite{Breitenlohner:1987dg} and follow-ups like \cite{Gaiotto:2007ag, Bergshoeff:2008be, Bossard:2009at, Mohaupt:2011aa} but not with the same degree of physical insight as offered by reference \cite{Heidenreich:2020upe}. What it does offer though is an immediate construction of multi-center solutions with arbitrary positions for the centers, nicely revealing the no-force property. The necessary condition is exactly that the product of surface gravity and horizon size vanishes.

\section{No forces and time-like reduction}
We briefly review the timelike reduction of 4d black holes following \cite{Breitenlohner:1987dg}. After that we sketch how this is generalised to black branes and black holes in other dimensions. 
\subsection{4d black holes}
We start with the general 2-derivative action for the space-time metric $g_{\mu\nu}$, Abelian gauge fields $B^I_{\mu}$ with $I,J,\ldots$ labeling the gauge field and scalar fields $\phi^r$ with $r,s,\ldots$ labeling the scalars: 
\begin{equation}\label{4Daction}
S_4=\int\Bigl(\tfrac{1}{2}\star R_4 - \tfrac{1}{2}G_{rs}\star\d
\phi^r\wedge\d\phi^s-\tfrac{1}{2}\mu_{IJ}\star G^I\wedge
G^J+\tfrac{1}{2}\nu_{IJ}G^I\wedge G^J\Bigr)\,.
\end{equation}
The fieldstrengths are $G^I=\d B^I$. The symbols $G_{rs}$, $\mu_{IJ}$, $\nu_{IJ}$ are symmetric matrices that depend on the scalars $\phi$. 
$G$ and $\mu$ are required to be positive definite. 

The Ansatz for black hole solutions can be written as
\begin{align}
& \d s_4^2 =-\e^{2U}(\d t + A_{KK})^2 + e^{-2U}\d s_3^2\,,\nonumber\\
& B^I= \tilde{B}^I + Z^I(\d t + A_{KK})\,,\label{reductionAnsatz}
\end{align}
where $\tilde{B}^I$ and $A_{KK}$ are vectors and $U$, $Z^I$ are
scalar fields on the spacelike slice with metric $ds^2_3$. When $A_{KK}$ can not be redefined away
($\d A_{KK}\neq 0$) there is a non-zero NUT charge and the solution is stationary but non-static. Note that we wrote the Ansatz like a Kaluza--Klein reduction over time, although time is non-periodic. In 3d the vectors $A_{KK}$ and $\tilde{B}^I$ can be dualised to scalars such that we end up with a sigma model coupled to scalars after compactification:
\be
S_3=\int\Bigl(\tfrac{1}{2}\star R_3 - \tfrac{1}{4}a_{ij}\star\d \Phi^i\wedge\d\Phi^j\Bigr)\,.
\ee
This sigma model has \emph{indefinite} metric $a_{ij}$ with the ``wrong sign'' scalars given by the electric  and  magnetic potentials. The explicit reduction and sigma model is presented in the appendix for completeness.

The Ansatz for the 3d metric with spherical symmetry is:
\begin{equation}\label{metric1}
\mathrm{d}s^2_3=\exp[4A(\tau)]\mathrm{d}\tau^2 +
\exp[2A(\tau)]\mathrm{d}\Omega_2^2\,,
\end{equation}
where $\d\Omega_2^2$ is the metric on the unit 2-sphere. The 3d scalar fields then depend on $\tau$ only, $\Phi^i=\Phi^i(\tau)$. 

Note that the Ansatz picks a certain gauge for the radial coordinate $\tau$. The reason for the gauge in (\ref{metric1}) is that the scalar equations of motion reduce to a geodesic problem with $\tau$ as an affine coordinate along the geodesic:
\be
S_{\Phi}  = -\tfrac{1}{4}\int \d\tau a_{ij}\dot{\Phi}^i\dot{\Phi^j} \,.
\ee
This means that the total velocity squared is a constant $c$
\be
\frac{1}{4}a_{ij}\dot{\Phi}^i\dot{\Phi^j} = c\,,
\ee
which can be positive, zero or negative. \emph{The crucial bit of this formalism is that the Einstein equations do not care about the details of the sigma model, they only see the constant $c$.}  The solution for the 3d metric reads
\begin{align}
& c = 0\quad  :e^{2A} = \frac{1}{\tau^2}\,.\\
& c > 0\quad  :e^{2A} = \frac{c}{\sinh^2(\sqrt{c}\tau)}\,.
\end{align}
The 3D metric with $c=0$ is just flat space in coordinates where $\tau=1/r$.
Solutions with $c<0$ have naked singularities when lifted to 4d and are thus unphysical (over extremal). But note that the 3d metrics with negative $c$ are smooth wormholes. It is the scalar $U$ that tends to develop singularities in the 4d metric after uplift. 

A crucial observation is that non-negative $c$ is related to the product of the temperature $T$ (surface gravity) and the entropy $S$ (horizon area) of the 4d black hole \cite{Ferrara:2008hwa}:
\be
c= 4 S^2T^2\,.
\ee
Most of the literature on the timelike reduction is concerned with solving the geodesic equations. This can be done exactly and explicitly for theories with more than 8 supercharges and for many interesting theories with 8 supercharges since then the sigma model will be some symmetric coset space with indefinite metric, whose geodesics are found using simple group theory. This is not our interest here. Our interest lies in the no-force condition derived in \cite{Heidenreich:2020upe} where solutions with $c=0$ were dubbed \emph{quasi-extremal}. To show that these solutions obey a no-force condition is now rather straightforward and is implied in the original work of \cite{Breitenlohner:1987dg}. Note that $\tau=1/r$ is an harmonic on the 3d metric. One can easily show that replacing $\tau$ by a generic multi-center harmonic 
\be
\frac{1}{r}\rightarrow \sum_i \frac{a_i}{|\vec{x}-\vec{x_i}|}\,,
\ee
still solves all 3d (and thus 4d) equations of motion if $c=0$. In the above equation $a_i$ are some arbitrary coefficients and $\vec{x_i}$ the position of the centers. The (3d) scalar equations of motion are satisfied as a consequence of the function being harmonic. This is sometimes called the harmonic map technique. To see that the 3d metric is still the same flat metric one simply observes that the 3d Einstein equation is 
\be
R_{ab} = \frac{1}{2}a_{ij}\partial_a\Phi^i\partial_b\Phi^j\,.
\ee
By using the chain rule one directly verifies that the right hand side vanishes if $\Phi^i$ describes a lightlike geodesic with the general harmonic as affine parameter. This is the generalisation of the familiar fact that extremal Reisnner-Nordstr\"om solutions can be put together without them feeling a force. Also these solutions can be obtained by writing the metric in terms of a harmonic function on 3d space that is multi-centered. 

We conclude that the spherical solutions with $c=0$ obey a no-force condition since they can replaced by multi-center ones. The crucial ingredient is the vanishing of the 3d Euclidean energy-momentum tensor ($c=0$) and that condition lifts to the product $ST$ being equal to zero in 4d, reproducing the findings of \cite{Heidenreich:2020upe}.
Note that this result would also apply to solutions with NUT charge .

\subsection{Black branes and other dimensions} 
The extension of the previous method where we map spherical black hole solutions to geodesic curves on the scalar manifold in 3d carries over to higher dimensions. Starting in $d+1$-dimensional space time and reducing over time one obtains a sigma model coupled to Euclidean gravity in $d$ dimensions:
\be\label{general}
S_d=\int\Bigl(\tfrac{1}{2}\star R_d - \tfrac{1}{4}a_{ij}\star\d \Phi^i\wedge\d\Phi^j\Bigr)\,, 
\ee
One difference with 4d spherical black holes is the absence of magnetic charges and that relates to the fact that vectors do not dualise to scalars in dimensions higher than 3. Similarly, the NUT-vector is put to zero since it would not be consistent with spherical symmetry. The Ansatz for the instanton geometry in $d$ dimensions is
\be
\d s^2 = e^{2(d-1)A}\d\tau^2 + e^{2A}\d \Omega_{d-1}^2\,, 
\ee
and is again chosen such that $\tau$ is an affine coordinate on the scalars. The solutions for non-negative $c$ are:
\begin{align}
& c = 0\quad  :e^{2A} = \frac{1}{\tau^{\frac{2}{d-2}}}\,.\\
& c > 0\quad  :e^{2A} = \beta^{\frac{2}{d-2}}\sinh^{-\frac{2}{d-2}}(\beta\tau)\,,\qquad \beta =\sqrt{\frac{2(d-2)}{d-1}c}  \,.
\end{align}
Again $c=0$ corresponds to a flat metric with $\tau$ being the spherical harmonic: $1/r^{d-2}$.
The negative $c$ solutions describe smooth Euclidean wormhole geometries with singular scalars in $d$ dimensions and lift to naked singularities in $d+1$ dimensions. Again one can verify that $c$ relates to a product of entropy and temperature.

Finally, we can consider black p-branes in $d+1$ dimensions. They are sourced electrically by $p+1$ forms. Since the spatial worldvolume is translation invariant we can formally reduce over the spacelike worldvolume to a black hole in $d-p+1$ dimensions where the electric $p+1$ forms reduce to vectors and scalars in $d+1$ dimensions. Once we mapped the problem to a black hole solution our previous results go through. A discussion of this can be found in section 2 of \cite{Bergshoeff:2008be}. A noteworthy difference is that now entropy becomes entropy density of the higher-dimensional brane.

\section{Axionic instantons}
In this section we shift gears and relate the previous precise results in classical general relativity to a more involved topic in semi-classical quantum gravity; the meaning of Euclidean axion wormholes introduced by Coleman \cite{Coleman:1988cy}, and more general, instanton solutions sourced by axion fields. Our discussion will be necessarily more heuristic and aims at providing more circumstantial evidence to the claim in \cite{Hertog:2018kbz} (and \cite{VanRiet:2020pcn}) that Euclidean wormholes of the Coleman type do not contribute to the path integral describing nucleation and absorption probabilities for baby universes carrying axion charges.

The technique of \cite{Breitenlohner:1987dg} explained in the previous section, relies on reducing a black hole in $d+1$ dimensions over its timelike Killing vector to a Euclidean ``instanton" solution in $d$ dimensions. Such timelike reductions are formal consistent truncations, not more. However, the formalism does shed light on discussions surrounding the notion of extremality for gravitational instantons sourced by axions \cite{Brown:2015iha, Heidenreich:2015nta, Andriolo:2020lul, VanRiet:2020pcn}.  

Instantons that can be related to changes in axion charges potentially contribute in path integrals with boundary conditions that have fixed axion charges, and hence fixed axion momenta. For such path integrals the effective action in Euclidean signature has a wrong sign for the axion kinetic term \cite{Burgess:1989da}. Just as it did for axions that are generated by dimensional reduction over time. So such instantons are again found by solving the system (\ref{general}) with indefinite sigma model metric. An example of the lightlike geodesic solution is the supersymmetric D-instanton in Euclidean IIB supergravity. The timelike geodesics, that is the wormholes, lifted to unphysical over-extremal solutions when they were found from timelike reduction. However, there is no timelike reduction in this context. And it has been shown that even the scalars can be entirely smooth even when embedded in Euclidean AdS space\cite{ArkaniHamed:2007js, Katmadas:2018ksp}. These wormholes are instead examples of Coleman's Euclidean wormholes that would corresponds to instanton processes that represent nucleation and absorption of baby universes that carry away or add axion charges \cite{Coleman:1988cy}. Are they physical? This is a long debated question which has been nicely reviewed in \cite{Hebecker:2018ofv}. If they are physical they do lead to puzzles in our understanding of quantum gravity and hence the easiest option to is claim they do not contribute in the path integral in line with general Swampland arguments regarding the Hilbert space of quantum gravity \cite{McNamara:2020uza}. 

If correct then both axion wormholes coming from timelike reductions and axion wormholes from ``a priori'' Euclidean theories are unphysical. Indeed, it was shown rather explicitly in \cite{Hertog:2018kbz} that such wormhole saddle points are unstable in the Euclidean sense and cannot contribute.  In fact, the instabilities found in \cite{Hertog:2018kbz} are in the non-homogenous sector and point to a defragmentation of the wormhole \cite{VanRiet:2020pcn}; the wormholes are ``self-repulsive" in a Euclidean sense. This self-repulsive property is the Euclidean analog of the self-repulsive particles in the Weak Gravity Conjecture context for particles and black holes. The explicit self-repulsive nature of wormholes can also be argued by evaluating the action of a probe BPS D-instanton  near a wormhole background \cite{VanRiet:2020pcn}. The conclusion is that fragmented wormholes have lower action. Hence the proper saddle points should be the ones where the fragmentation is taken to its limit and the wormhole neck becomes of Planckian size where a semi-classical treatment is impossible. These latter objects are more likely to be physical and contribute to the path integral, consistent with the fact that axion shift symmetries are broken by a potential generated through instantons \cite{Hebecker:2018ofv}. This is also consistent with the stability analysis of \cite{Hertog:2018kbz} since the modes that lower the action are concentrated around the neck and disappear in the limit of small wormholes.

A final confirmation of the spurious nature of Euclidean axion wormholes comes from AdS/CFT, as suggested first in \cite{ArkaniHamed:2007js}. Since extremal instantons are dual to (anti-) self-dual Yang-Mills instantons ($\star F=\pm F$), one does not expect ``over-extremal" configurations in the dual, since (anti-) self-dual configurations have the lowest action for a given winding number (axion charge). Indeed the dual one-point function $\langle \text{Tr}(\star F \pm F)^2 \rangle $ for Euclidean wormholes is negative, signalling an inconsistency \cite{Katmadas:2018ksp}. However, this argument would also apply to the single-charge microscopically sized wormholes. Derivative corrections could be significant in the bulk but not near the boundary where the one-point functions are computed. It remains an excellent research question to sort out how the violation of the positivity bound is consistent with microscopic wormholes contributing to the path integral. 

So we have argued that the self-repulsive \emph{macroscopic} instanton is unphysical whereas the self-repulsive \emph{microscopic} one is likely to be physical. Therefore, the analogy between particles, black holes and microscopic, macroscopic instantons seems to work.

\section*{Acknowledgments}
I like to thank Stefano Andriolo for discussions and Ben Heidenreich for useful correspondence. This work is supported by the KU Leuven C1 grant ZKD1118C16/16/005. 

\appendix
\section{Dimensional reduction over time}
The dimensional reduction of the action (\ref{4Daction}) gives:
\begin{align} S_3&=\int\Bigl(\tfrac{1}{2}\star R_3 - \star\d
U\wedge\d U +\tfrac{1}{4}\e^{4U}\star F_{KK}\wedge F_{KK} -
\tfrac{1}{2}G_{rs}\star\d \phi^r\wedge\d\phi^s \nonumber \\ & +
\tfrac{1}{2}\mu_{IJ}\e^{-2U}\star \d Z^I\wedge \d Z^J
-\tfrac{1}{2}\e^{2U}\mu_{IJ}\star(\tilde{G}^I+Z^IF_{KK})\wedge
(\tilde{G}^J+Z^JF_{KK})\nonumber
\\ & -\nu_{IJ}(\tilde{G}^I+Z^IF_{KK})\wedge \d Z^J\Bigr)\,,
\end{align}
where $\tilde{G}^I=\d\tilde{B}^I$, $F_{KK}=\d A_{KK}$. As usual the
vectors $A_{KK}$ and $\tilde{B}^I$ can be dualised to scalars $\chi$
and $Z_I$ by adding the following Lagrange multipliers to the action
\begin{equation}
S'_3=S_3 +\chi \d F_{KK} + Z_I\d \tilde{G}^I\,.
\end{equation}
Varying the action $S_3'$ with respect to $F_{KK}$ and $\tilde{G}^I$
gives the equations of motion
\begin{align}
\d Z_J  &=-\e^{2U}\star \mu_{IJ}(\tilde{G}^I+Z^IF_{KK})-\nu_{IJ}\d Z^I \,,\label{dual1}\\
\d \chi &= \tfrac{1}{2}\e^{4U}\star F_{KK} +Z^I\d
Z_I\,.\label{dual2}
\end{align}
Dualisation of the action $S_3$ is obtained by eliminating $F_{KK}$
and $\tilde{G}^I$ from  the action $S_3'$ using (\ref{dual1},
\ref{dual2}). If we furthermore define, $2\chi \equiv a+Z^IZ_I$ we
find
\begin{align}
S_3&=\int\Bigl(\tfrac{1}{2}\star R_3 - \star\d U\wedge\d U
-\tfrac{1}{2}G_{rs}\star\d \phi^r\wedge\d\phi^s +\tfrac{1}{2}\e^{-2U}\star \d {\bf Z}^T \wedge \mathcal{M}_4 \d {\bf Z} \nonumber\\
& -\tfrac{1}{4}\e^{-4U}\star(\d a + {\bf Z}^T \mathbb{C}\d {\bf
	Z})\wedge (\d a + {\bf Z}^T \mathbb{C}\d {\bf
	Z})\Bigr)\,,\label{sigma3D}
\end{align}
where
\begin{equation}
{\bf Z}\equiv (Z^I, Z_I)\,,\qquad
\mathbb{C}=\begin{pmatrix}0 & -1\\
+1 & 0 \end{pmatrix}\,,\qquad \mathcal{M}_4 =
\begin{pmatrix} \mu
+\nu\mu^{-1}\nu & \nu\mu^{-1}\\
\mu^{-1}\nu & \mu^{-1}
\end{pmatrix}\,.
\end{equation}
This is an action of gravity coupled to a sigma model for the 3D scalars $U,\phi^r, Z^I, Z_I, a$ where $U$ will correspond to the 4D warpfactor, $Z^I$ to electric potentials and $Z_I$ to magnetic potentials. One can show that it is consistent to construct solutions with zero NUT charge for spherical solutions and then the second line in equation (\ref{sigma3D}) can be dropped all together and the scalar $a$ is then eliminated. Then the scalars $Z^I,Z_I$ appear shift symmetric in the action (\ref{sigma3D}). 
\small{
\bibliography{refs}}
\bibliographystyle{utphys}

\end{document}